TITLE: Respiratory rhythm entrains membrane potential and spiking of non-olfactory neurons

RUNNING TITLE: Respiratory modulation in the brain: deep in the cells

AUTHORS: Maxime Juventin[1], Mickael Zbili[1,2], Nicolas Fourcaud-Trocmé[1], Samuel Garcia[1], Nathalie Buonviso[1]* and Corine Amat[1]*

AFFILITATION:

[1] Lyon Neuroscience Research Center (CRNL) Inserm U 1028, CNRS UMR 5292, University Lyon1, 69675 Bron, France

[2] Blue Brain Project, École Polytechnique Fédérale de Lausanne (EPFL), Campus Biotech, Geneva, Switzerland

*These authors jointly supervised this work.

CORRESPONDING AUTHOR: Corine Amat, Lyon Neuroscience Research Center (CRNL) Inserm U 1028, CNRS UMR 5292, University Lyon1, 69675 Bron, France, corine.amat@univ-lyon1.fr

ABSTRACT

In recent years, several studies have tended to show a respiratory drive in numerous brain areas so that the respiratory rhythm could be considered as a master clock promoting communication between distant brain areas. However, outside of the olfactory system it is not known if respiration-related oscillation (RRo) could exist in the membrane potential (MP) of neurons neither if it can structure spiking discharge. To fill this gap, we co-recorded MP and LFP activities in different non-olfactory brain areas: median prefrontal cortex (mPFC), primary somatosensory cortex (S1), primary visual cortex (V1), and hippocampus (HPC), in urethane-anesthetized rats. Using respiratory cycle by respiratory cycle analysis, we observed that respiration could modulate both MP and spiking discharges in all recorded areas. Further quantifications revealed RRo episodes were transient in most neurons (5 consecutive cycles in average). RRo development in MP was largely influenced by the presence of respiratory modulation in the LFP. Finally, moderate hyperpolarization reduced RRo occurence within cells of mpFC and S1. By showing the respiratory rhythm influenced brain activity deep to the MP of non-olfactory neurons, our data support the idea respiratory rhythm could mediate long-range communication.

INTRODUCTION

In recent years, a resurgence of publications has proposed respiration as a master clock for brain rhythms, following many observations that reported the breathing rhythm as being able to modulate complex behaviors or cognitive process (Heck et al. 2017, 2019; Maric et al. 2020; Zaccaro et al. 2018). The exploratory behavior of rodents, consisting in rhythmical sampling of the environment, is a perfect example since the onset of each respiratory cycle initiates a "snapshot" of the orofacial sensory environment, comprising sniffing, whisking and head movements (Cao et al. 2012; Kurnikova et al. 2017; Moore et al. 2013). Similarly, licking synchronizes with breathing (Lu et al. 2013; Welzl and Bureš 1977). In humans, respiration modulates memory consolidation (Arshamian et al. 2018), associative learning (Waselius et al. 2019), cognitive performances (Nakamura et al. 2018; Perl et al. 2019; Zelano et al. 2016), sensory perception(Iwabe et al. 2014), and motor behaviors (Li et al. 2012; Li and Laskin 2006). Respiratory phase influences the initiation of a voluntary action (Park et al. 2020) or a cognitive task and its performance (Perl et al. 2019). All these data therefore suggest that the brain could have at least one processing mode depending on the respiratory phase.

Parallel to this literature on the effects of breathing on behavior has appeared an equally important literature on a respiratory drive of brain activity, both in human and rodent. In human, intracerebral EEG-breathing coherence has been observed notably in the hippocampus, amygdala, insula and the parietal lobe (Herrero et al. 2018; Zelano et al. 2016) and inhalation phase is correlated with an increase in the power of delta oscillations (Zelano et al. 2016). In rodent, a respiration-locked local field potential (LFP) oscillation has been described in the hippocampus (Yanovsky et al. 2014), the somatosensory cortex (Ito et al. 2014), the prefrontal cortex (Biskamp et al. 2017; Köszeghy et al. 2018), neocortical regions up to visual cortex and even subcortical areas (Girin et al. 2021; Tort et al. 2018). In several brain regions, respiration also modulates other rhythms. In the hippocampus, the timing of hippocampal slow-wave ripples is respiration-coupled (Liu et al. 2017). In the somatosensory cortex, delta oscillations are respiration-locked while gamma power is respiration-modulated (Ito et al. 2014). In the prefrontal cortex, fast gamma LFP oscillations (Zhong et al. 2017) and unitary activities (Biskamp et al. 2017) are phase-coupled with respiration.

This respiratory drive, described in such a large cerebral network, is at the moment assumed to arise in the activation of the olfactory bulb (OB) *via* mechanic stimulation of olfactory receptor cells by nasal respiratory airflows (Grosmaitre et al. 2007). Indeed, respiration-brain activity coupling dissipates in human when subjects breathe through the mouth (Zelano et al. 2016) or in rodents when OB is silenced (Ito et al. 2014; Liu et al. 2017; Lockmann et al. 2016; Yanovsky et al. 2014). This respiration-related activity is transmitted to the OB where it has been largely described (For review, Buonviso et al. 2006): respiration modulates both LFP and individual cells activities, including spiking discharge and membrane potential dynamics. In LFP signal, fast rhythms occurrence is shaped by the slow respiration-related oscillation (RRo), beta and gamma waves alternating on expiratory and inspiratory phases, respectively (Buonviso et al. 2003; Cenier et al. 2009). Mitral/tufted cells spiking discharge is strongly patterned by respiratory rhythm both in anesthetized (Briffaud et al. 2012; Cenier et al. 2009; Courtiol et al. 2011a; David et al. 2009) and awake rodent (Cury and Uchida 2010; Gschwend et al. 2012). Importantly, even the membrane potential of mitral/tufted cells is impacted by respiration(Ackels et al. 2020; Briffaud et al. 2012; Fourcaud-Trocmé et al. 2018). Particularly, RRo of membrane potential constrain fast oscillations to occur concomitantly with LFP signal, hence favoring interaction between fast LFP and spiking (Fourcaud-Trocmé et al. 2018). We and others have demonstrated that most of OB respiratory modulation is related to the amplitude of the inspiratory airflow (Courtiol et al. 2011a, 2011b; Esclassan et al. 2012; Oka et al. 2009). Respiratory influence is also present, to a lesser extent, in the olfactory cortex at the LFP and cellular levels (Piriform cortex: Courtiol et al. 2019; Litaudon et al. 2003, 2008;

Miura et al. 2012; Olfactory tubercle: Carlson et al. 2014). Directly in contact with nasal respiratory airflows, the olfactory system is thus strongly impacted by respiration dynamics.

Oppositely, while LFP activity in non-olfactory regions has been described as respiration-modulated, very little is known about respiratory modulation of individual cell activities. Only a few reports described spiking discharge modulated by respiration (Biskamp et al. 2017; Köszeghy et al. 2018). To our knowledge, only a BioRxiv preprint reported intracellular recordings in parietal cortex (Jung et al. 2019). There are no studies showing that the membrane potential (MP) of cells in non-olfactory regions can oscillate with respiration nor, as we have shown in mitral/tufted cells of the OB, that this slow respiration-related oscillation of MP can structure spiking discharge. Here we asked to what extent respiratory modulation could shape cellular activity, including MP and spiking activity, in non-olfactory areas. To this purpose, we co-recorded membrane potential and LFP activities in widespread brain areas, namely the median prefrontal cortex (mPFC), the primary somatosensory cortex (S1), primary visual cortex (V1), and hippocampus (HPC), in urethane-anesthetized rats. Using respiratory cycle by respiratory cycle analysis, we tracked respiratory modulation of cellular activities. We observed that both MP and spiking discharges could be respiration-modulated in all recorded areas, but more transiently than what we previously described in the OB (Briffaud et al. 2012). Particularly, respiratory modulation of MP was highly variable both between structures and between cells, very transient for most of cells and largely influenced by LFP state. Moreover, even if internal excitability state had a weak influence on the proportion of respiration-modulated cells at the whole population level, moderate hyperpolarization of a cell in mPFC and S1 reduced the strength of its modulation.

MATERIALS AND METHODS

Experimental procedures

*Animal care*

The 25 male Wistar rats (325.2±32.1g, Charles River) used were housed in groups of four in a temperature (22±1 °C) and humidity (55±10%) controlled room and exposed to a 12/12 h light/dark cycle (light onset, 6:00 am). Experiments were conducted during the light period (between 9:00 am and 3:00 pm). Food and water were available *ad libitum*. All experiments were carried out in accordance with Directive 2010/63/EU of the European Parliament and of the Council of the European Union regarding the protection of animals used for scientific purposes and in compliance with the ARRIVE guidelines. The experimental protocols were approved by the National Ethics Committee "Animal Experimentation Committee of Univ. Claude Bernard Lyon 1—CEEA-55" (Agreement APAFIS #17088).

*Animal preparation*

Rats were anesthetized with an intraperitoneal injection of urethane (1.5 g/kg), then, placed in a stereotaxic apparatus. Respiration was recorded with a homemade bidirectional airflow sensor placed at the entrance of the right nostril. Positive and negative flows corresponded respectively to inspirations and expirations. Body temperature was kept at 37 °C using a heating pad (Harvard Apparatus, Holliston, MA USA). A craniotomy of about 3x3 mm was performed above the site of recording: median prefrontal cortex, primary visual cortex, somatosensory barrel cortex or hippocampus. Dura was removed, and Ringer's lactate solution was regularly applied onto the brain to prevent from drying. At the end of the experiments, rats were euthanized by intracardiac injection of Dolethal©.

*In vivo Electrophysiological Recordings*

We made simultaneous intra- and extracellular recordings in freely breathing anesthetized rats. Intracellular recordings (n = 69 cells) were performed in four different areas of the left hemisphere: the medial prefrontal cortex (mPFC, AP: 3.1±0.4 mm, MD: 0.7±0.2 mm), the primary somatosensory cortex (S1, AP: -2.4±0.4 mm, MD: 5.0±0.1 mm), the primary visual cortex (V1, AP: -5.3±0.2 mm, MD : 4.5±0.8 mm) and the hippocampus (HPC, -3.5±0.7 mm, MD : 2.6±0.7 mm). Borosilicate glass capillaries (o.d. = 1.5 mm; i.d. = 0.86 mm, Harvard Apparatus, Holliston, MA, USA) were pulled with a horizontal puller (model P-97, Sutter Instruments, Novato, CA, USA). The micropipettes were filled with a 2 M potassium acetate solution, and their resistances ranged from 50 to 120 MΩ. Microelectrodes were lowered using a piezo manipulator (PM-10, World Precision Instruments, Sarasota, FL, USA). The electrophysiological signal was amplified and low-pass filtered at 10 kHz by an intracellular amplifier (Axoclamp 2B, Axon Instruments, Foster City, CA, USA). The signal was then digitalized at 25 kHz (NI USB-6211, National Instruments, Austin, TX, USA) and stored on a computer using Neurolabscope, a homemade software.

*Internal polarity state*

Internal polarity states (studied in Figure 4) were visually encoded based on sign (positive or negative) and intensity of direct current (DC) injection, amplitude of MP variation and firing activity. We identified 4 states: strong hyperpolarization (Hyp2), light hyperpolarization (Hyp1), basal level (Base), depolarization (Dep). A cell was not systematically recorded under each of the 4 states. *A posteriori* verification comforted our classification. Whatever the structure (Figure 4A), basal level (Base) represented the state where none (N = 42 cells) or a very low stabilizing current (-0.07±0.06 nA, N = 14 cells) was injected. Hyp1 level corresponded to a MP hyperpolarization around 7 mV from basal level at which the spike discharge decreased by 60%. This level was achieved via a -0.28±0.04 nA (N = 29 cells) DC injection. Hyp2 level corresponded to a MP hyperpolarization around 12 mV from basal level at which the spike discharge decreased by 90% to 100%. This level was achieved via a -0.41±0.04 nA (N = 36 cells) DC injection. Dep level corresponded to a MP depolarization around 7 mV from basal level at which the spike discharge increased by at least 10%. This level was achieved via a 0.17±0.04 nA (N = 22 cells) DC injection. For the cell-by-cell analysis (Figure 4B Base-Hyp1), to limit influence of external factors, we selected pair of recordings close in time. In mPFC, cells (N = 9 cells) received -0.17±0.02 nA for Hyp1 which resulted in a hyperpolarization of 7.6±2.5 mV and a spike discharge decrease of 74.4±7.5 %. Latency between end of Base recording and beginning of Hyp1 recording was 3.2±1 min. In S1, cells (N = 4 cells) received -0.19±0.06 nA for Hyp1 which resulted in a hyperpolarization of 8.4±5.2 mV and a spike discharge decrease of 50.0±26.6 %. Latency between end of Base recording and beginning of Hyp1 recording was 3.3±1.4 min. In V1, cells (N = 3 cells) received -0.22±0.06 nA for Hyp1 which resulted in a hyperpolarization of 13.9±7.6 mV and a spike discharge decrease of 55.2±13.1 %. Latency between end of Base recording and beginning of Hyp1 recordings was 2.3±0.8 min. In HPC, cells (N = 3 cells) received -0.26±0.05 nA for Hyp1 which resulted in a hyperpolarization of 7.6±3.0 mV and a spike discharge decrease of 50.5±10.0 %. Latency between end of Base and beginning of Hyp1 recordings was 1.3±0.2 min.

Local field potential (LFP) was recorded simultaneously with all intracellular recordings using silicon probes (Neuronexus Technology, Ann Arbor, MI, USA) placed as close as possible to the glass microelectrode. The signal was amplified, and low-filtered at 5 kHz with a homemade amplifier.

Data analysis

All following analyses have been made using homemade Python scripts.

*Respiratory signal processing*

Respiratory cycles were detected as previously described in Roux et al. 2006. Briefly, respiratory signal was high-pass filtered to reduce noise, then zero-crossing points were detected to delineate inspiration and expiration start points. Finally, time of each respiratory cycle was linearly converted to phases ranging from 0 to 1. To summarize, all respiratory cycles were detected, inspiration and expiration phases were well determined, and respiratory phases were computed.

*Rescaling time basis of signals with respiratory phase*

We wanted to track variations of signals of interest (LFP, MP, AP discharge) related to respiratory rhythm. However, respiratory cycles displayed duration variability, and inspiration-expiration ratio from two cycles of same duration could differ. Thus, the respiratory phase was the appropriate time scale to reduce jitter due to respiratory cycle variability. Similarly to Roux et al., 2006., we rescaled signals of interest as a function of respiratory phase. Briefly, after having detected respiratory cycles, signal during each cycle was interpolated into a 2000 points vector. To account for the asymmetry of respiratory cycles, the mean inspiration ratio (inspiration duration/cycle duration) was computed for each recording. Cycle by cycle interpolation preserved the ratio, *i.e.,* for a 0.33 ratio, interpolation attributed 33% of the 2000 points to the inspiration and the rest to the expiration. This rescaling allowed the construction of respiration triggered MP average (Figure 1C bottom) or respiration triggered AP histogram (Figure 1C top). We made one exception to construct Figure 3B as we wanted to compare the preferred phase of depolarization between cells, we rescaled all recordings with the same ratio 0.4 in order to align all cell respiratory phases.

*Time dynamics of respiration-related MP oscillations*

The method described hereafter aimed at determining if a given signal followed the respiratory rhythm on a cycle-by-cycle basis. It was applied to AP-removed membrane potential and LFP signals. To remove APs, single APs were cut from 4ms before to 5 ms after the peak. Burst were cut from 30ms before the first AP to 30 ms after the last AP. Cut signal was replaced with a linear interpolation between the two extremities. Several cutting lengths were tested. Visually satisfying parameters were kept.

First step: Identification of local oscillation related to respiration. Signal of interest was rescaled towards respiratory phase as explained previously. As we wanted to follow the time dynamics of respiratory oscillations, we scanned recordings with windows of 4 respiratory cycles, with an overlap of 3 cycles between consecutive windows. In each window we computed respiration triggered median signal, which amplitude should be high when signal oscillations are synchronized to respiration. To question the existence of such a local respiratory related oscillation, we compared median signal waveform amplitude with surrogate amplitudes (N = 500). One surrogate of one window was built as follow: we randomly created 2 pairs of cycle (for example cycles 1&3, and 4&2). For one pair we phase shifted with a random ($\varphi_1$) phase the first cycle. The other cycle of the pair was phase shifted with the opposed phase ($\varphi_1+\pi$). Second pair underwent the same transformation, with another random phase-shift ($\varphi_2$). By doing so, within each pair (phase-opposing) and between pairs (random phase-shifting) respiratory influence was neutralized when averaging, thus surrogate data contained only the signal part unrelated to respiration. We considered a local waveform was respiratory related if window median signal amplitude was larger than 95$^{th}$ percentile of surrogate amplitudes. This algorithm allowed to classify local signal in each window as respiratory-related or not.

Second step: Refinement from local oscillation to individual cycles. To reduce false-detection and to increase time precision, we wanted this encoding to be realized cycle by cycle. From the windows previously identified as respiratory related windows, we wanted to extract the cycles which really contributed to this respiratory waveform. To ensure that individual cycles within the window looked like the local waveform, we computed for each cycle the scalar product between the cycle signal and the window median signal. Since we focused on waveform, not on amplitude, we z-scored the signals before performing the scalar product ((X-mean(X))/std(X)). We assessed significance by comparing the result to scalar products between the cycle signal and 500 surrogated window signals (we kept the previous surrogates). We considered the individual cycle sufficiently similar to the local waveform if the scalar product value was greater than the 95$^{th}$ percentile of surrogated scalar product values. If so, score of this cycle was incremented by 1. Since a cycle may be scanned up to four times (in four consecutive windows), score of individual cycle ranged from 0 to 4. We classified cycles in three respiratory related categories as follow: 0 and 1: non-modulated cycles, 2: undetermined, 3 and 4: modulated cycles. Finally, we eliminated all single or doublet cycles identified as respiratory entrained, and they were transferred to the undetermined category. Indeed, we considered that respiratory related events of strictly less than 3 consecutive cycles were not physiologically relevant and too much susceptible to false detection.

For each cell we were then able to compute the probability of respiratory entrainment (expressed in percent), defined as the ratio between the number of respiration-related cycles and the total number of cycles. For further analysis, we considered a cell to be sensitive to respiration if respiratory entrainment was strictly higher than 2.5%. This threshold avoided overestimation of respiration-sensitive cells for long recording. Overall, the method has been designed to reduce false detection to the minimum; therefore, we may underestimate the respiratory entrainment.

*Respiration-triggered membrane potential*

Using the classification method described in the previous paragraph, we could isolate membrane potential cycles related to respiration. These cycles were then stacked, and we computed the median signal. For each respiration-sensitive cell, the amplitude of respiration-related oscillations was defined as the maximum/minimum difference of this median signal. For the 4 examples in Figure 1, 95% confidence intervals represent the sum of standard error (1.91*SEM) at the minimum and the maximum points. To analyze preferred phase of the depolarization, we computed the median respiration-triggered signal, then we extracted the mean vector and finally we kept its angle. For this latter analysis we only used respiration sensitive cells whose signal contained strictly more than 5 respiratory-related cycles.

*Respiration-triggered spiking pattern*

AP times were rescaled to a respiratory cycle phase basis. Therefore, AP were identified by the number of the respiratory cycle they belong to, and their respiratory phase within that cycle. Once respiratory cycles were classified in respiratory-related categories, we could sort AP according to these categories. Then, for each cell and each respiratory-related category we built respiration triggered discharge histogram. For each cell, three experts, independently, visually encoded spiking patterns associated to each respiratory-related category. Spiking patterns were classified as: respiration driven or not, and with not enough spikes to be determined. Classification was inspired from Buonviso et al. 2006.

*Power spectrum*

For the 4 examples in Figure 1, Morlet wavelet time-frequency maps (0.4-6Hz) for respiratory signal and AP-removed membrane potential were computed for whole recordings. Then we extracted the 4s time windows showed in the figure and computed the mean across time by frequency to estimate the power spectrum. Spectra were normalized by their sum to get a density.

*Extracellular signal (LFP) processing*

LFP signal was 0.3Hz high pass filtered. Then, similarly to membrane potential oscillations, LFP-respiration relationship was classified cycle by cycle. Using this classification, the relationship to respiratory rhythm was then compared between MP and LFP.

*Conditional probabilities*

In Figure 5D, we looked at the conditional probability to observe MP RRo relative to LFP dynamics. In other words, we split recording according to LFP RRo encoding. On the one hand we computed MP RRo probability within cycles identified as LFP RRo (without % cycles threshold), and on the other hand we still computed the MP RRo probability but this time within cycles identified as LFP activity not related to respiration (other LFP activity). Figure 5E is the complementary view since we computed the conditional probability to observe LFP RRO according to MP dynamics (RRo or other activity).

*Statistics*

Statistics were made with Scipy (version 1.4.1) python package (scipy.stats.wilcoxon, scipy.stats.mannwithneyu, scipy.stats.kruskal). All averages are presented under the format mean±SEM. If not, it is explicitly mentioned.

RESULTS

**Evidence for a respiratory modulation of cell membrane potential in non-olfactory structures**

*In vivo* intracellular activities recordings were obtained from 4 non-olfactory structures in 25 freely breathing anesthetized rats; specifically, 34 cells were recorded in median prefrontal cortex (mPFC, N = 9 rats), 13 cells in somatosensory cortex (S1, N = 4 rats), 5 cells in visual cortex (V1, N = 4 rats) and, 17 cells in hippocampus (HPC, N = 8 rats). The respiratory frequency ranged from 1.44 Hz to 2.08 Hz (1.78±0.03 Hz, N = 25 rats).

In all these four areas, we found cells with a respiration-related oscillation (RRo) of their membrane potential (MP) and a respiration-related (RR) spiking discharge, as shown on the raw data examples (Figure 1A). Indeed, the filtered respiratory signal (light gray) fits the raw traces of the intracellular MP oscillations. In Figure 1B, the power spectral density analysis of the MP signal (black PSD) exhibits a clear power peak at the respiratory frequency (grey PSD) which confirms that MP follows respiratory rhythm in these examples.

In order to properly characterize the relation between MP and respiratory rhythm, we developed a cycle-by-cycle method for detecting the respiratory cycles at which MP oscillated at the respiratory rhythm (see materials and methods for details). We plotted the median of respiration-triggered membrane potential (Figure 1C bottom panels) and the histograms of spiking occurrence during these cycles (Figure 1C top panels). In the examples of Figure 1, MP RRo are large (median amplitude ± 95% confidence interval: mPFC = 5.5±0.9 mV, S1 = 14.5±2.1 mV, V1 = 8.8±2.2 mV, HPC = 3.1±0.8 mV; Figure 1C bottom panels), and APs occurrences are clearly related to respiratory cycle (top panels in Figure 1C). Moreover, on these examples, MP RRo and firing pattern are similarly phase-locked to respiration, suggesting a strong correlation between MP RRo and the associated firing. Indeed, spikes appeared on the top of the oscillation, where the MP is the most depolarized.

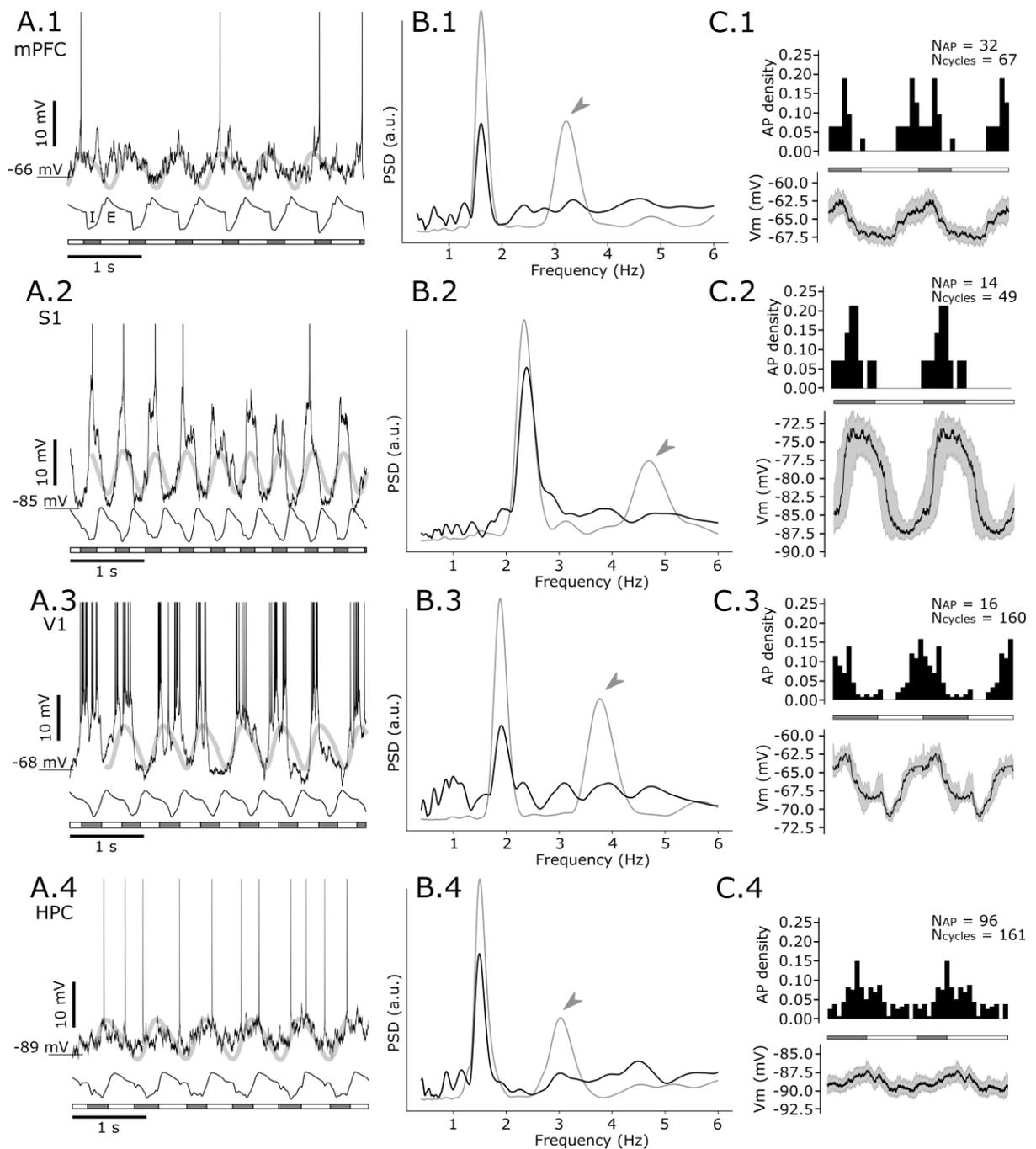

**Figure 1 : Intracellular activity of widespread brain areas correlates with respiratory rhythm.** A. Raw traces of intracellular activity from mPFC, S1, V1 and HPC (respectively subpanels 1, 2, 3, 4). A panels: Top trace: intracellular raw trace in black, overlaid with a filtered (3 Hz low pass) respiratory signal, in grey, which has been time-shifted to highlight

MP RRo. Mid trace: Respiratory signal. Bottom: Respiratory phase, with grey bars representing inspiration phases, white bars representing expiration phases. Time scale bar = 1s, amplitude scale bar = 10 mV. E = expiration, I = inspiration. AP have been truncated. B panels: MP (black) and respiratory signal (grey) power density spectrum over the 4s of signal presented in A. Arrow heads highlight 1$^{st}$ harmonic of the spectrum. C panels Respiration-triggered AP firing histogram (top) and respiration triggered MP (bottom). Data of respiration-triggered figures are extracted from several cycles during which MP is identified as correlated with respiration. Cycles and AP numbers are indicated on the right top. To visually reinforce oscillatory behavior, plots have been duplicated. Respiratory phase is reminded in between the two respiration-triggered figures. Grey and white bars correspond respectively to inspiration and expiration. Bottom: respiration triggered MP where black line corresponds to the median, shaded areas delineate the 25$^{th}$ and 75$^{th}$ percentiles.

Thus, we evidenced that a respiratory modulation can be expressed in the intracellular activity (MP and spiking) of cells in the four non-olfactory areas we recorded. The next step was to better characterize this MP respiratory modulation in each structure.

We first quantified the percentage of cells exhibiting MP RRo at rest. As explained in Methods, we defined an MP as being respiration-entrained only if we observed at least 3 consecutive RRo cycles. The first step was to measure the percentage of respiratory cycles eliciting an MP RRo in each cell for the 4 structures (Figure 2A, MP RRo cycles, MP No RRo cycles and MP which is at the limit to be classified as RRo respectively black, white and grey bars, see methods), then to draw the cumulative distribution of RRo MP cycles percentage across cells for each structure (Figure 2B). From these representations, we chose to classify a cell as respiratory entrained if it expressed at least 2.5% of RRo cycles (red lines in Figure 2A1-4). Under such conditions, respiration-entrained cells represented 53.1% of cells (17/32) in mPFC, 60% (6/10) in S1, 50% (2/4) in V1, and 80% (8/10) in HPC. Therefore, the three cortical structures (mPFC, S1 and V1) presented equivalent proportions of respiration-entrained cells while HPC presented a higher proportion. More specifically, the cumulative percentage curves (Figure 2B) show that 20% of the cells in S1 and mPFC had about 20% of their MP cycles modulated by respiration, more than 20% of HPC cells had about 75% of their cycles modulated, and no cell in V1 had more than 25% modulated cycles.

As seen in Figure 2A1-4, the percentages of MP RRo cycles, varied greatly from cell to cell in a same structure and between structures. Indeed, if we take into account only the cells exhibiting more than 2.5% of cycles with RRo MP (above horizontal lines), the percentage of RRo MP cycles ranged from 2.7% to 90.5% in mPFC (18.6±5.4%, N = 17 cells) and from 3.2% to 88.9% in HPC (29.5±11.4%, N = 8 cells), while it only extended from 6.5 to 33.3% in S1 (15.4±3.6%, N = 6 cells) and from 17.2 to 25.8% in V1 (21.5±3.1%, N = 2 cells) (Figure 2C). Note that three neurons (one in mPFC and two in HPC) exhibited an extreme proportion of MP RRo cycles (more than 90%). The low percentage of MP RRo cycles for some cells is certainly due to our highly selective method for detecting RRo cycles (see material and methods). Due to the great variability between cells, no significant difference in proportion of MP RRo cycles between structures was observed (Figure 2C, Kruskal Wallis test, p = 0.70, H(3) = 1.45).

To go further, we quantified for each entrained cell at resting MP (Base) the average duration of respiration-entrained episodes (*i.e.*, the number of consecutive cycles that presented MP RRo). As shown in Figure 2D, there was no significant difference in the duration of respiration-entrained episodes between structures (Figure 2D, Kruskal Wallis test, p = 0.18, H(3) = 4.77). In fact, the average number of consecutive MP RRo cycles was around five regardless of the structure (mPFC: 5.2±0.3, N = 17 cells; S1: 4.9±0.6, N = 6 cells; V1: 4.8±0.4, N = 2 cells; HPC: 6.0±0.6, N = 8 cells).

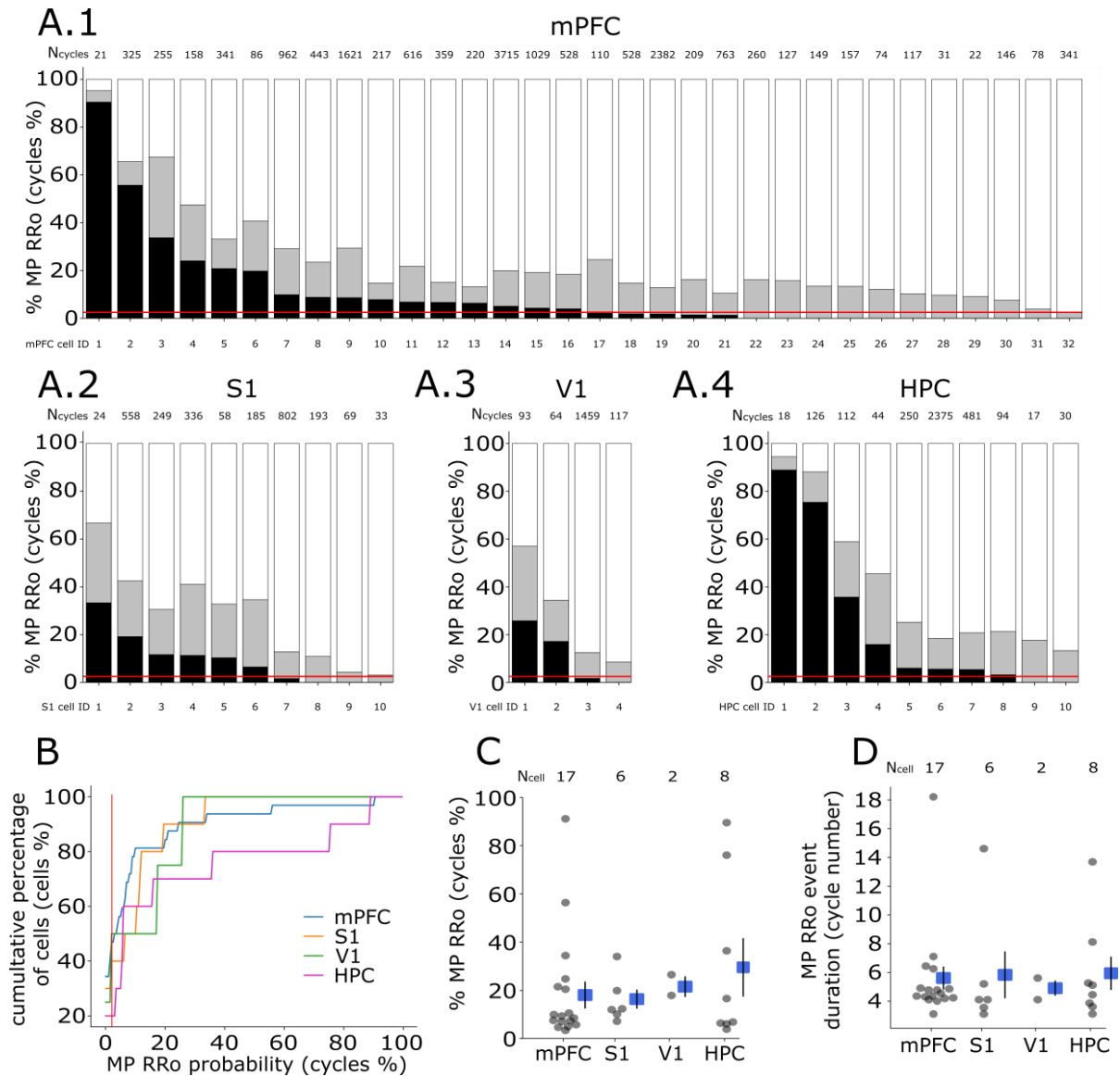

**Figure 2 : A significant number of cells in each structure exhibit MP RRo, yet individual episodes are short.** Only recordings at basal MP are considered. A. Distribution of MP encoding by cell in mPFC, S1, V1 and HPC (respectively subpanels 1, 2, 3, 4). Black, grey and white, bars correspond to the probability of RRo cycles, undetermined cycles and no RRo cycles respectively. Red lines indicate our 2.5% threshold. On the top of each bar is indicated the total number of recorded respiratory cycles. B. Cumulative distribution of RRo percentages

of each structure. Blue, orange, green and pink curves correspond to mPFC, S1, V1 and HPC respectively. Red lines highlight our 2.5% threshold. C. MP RRo probabilities across structures. Grey points represent MP RRo probability for individual cells, blue squares represent the mean by structure and black line represent SEM. Only cells with RRo probability above 2.5% are considered. No significant difference is observed (Kruskal-Wallis test, p = 0.69, H(3) = 1.45, $N_{mPFC}$ = 17 , $N_{S1}$ = 6, $N_{V1}$ = 2, $N_{HPC}$ = 8). D. MP RRo episode duration across structures. Grey points represent mean MP RRo episode duration of individual cells, blue squares represent the mean by structure and black lines represent SEM. The same cell pool as in B is used. No significant difference is observed (Kruskal-Wallis test, p = 0.80, H(3) = 0.98, $N_{mPFC}$ = 17 , $N_{S1}$ = 6, $N_{V1}$ = 2, $N_{HPC}$ = 8).

Taken together, these results show that at least 50% of cells of each recorded structures exhibited MP RRo. However, this respiratory entrainment of MP was 1) highly variable both between structures and between cells, 2) very transient for most cells (with mean episode duration of 5 cycles).

Then, to further explore the intrinsic dynamics of these RRo events, we studied both their amplitudes and phases. We first measured the amplitude of the median respiration-triggered MP (see material and methods for details) for each entrained cell (Figure 3A). In order to reduce the noise in the median respiration-triggered MP, we restricted the analysis to cells with at least 5 RRo cycles (Figure 3A, B). This revealed that amplitude of MP RRo was higher in S1 (11.2±2.5 mV, N = 6 cells) compared to mPFC (3.5±0.4 mV, N = 16 cells), HPC (3.7±0.5 mV, N = 7 cells), and V1 (5.6±1.5 mV, N = 2 cells) (Figure 3A). Statistical threshold was not reached except for S1 compared to mPFC and HPC (Kruskal Wallis test, p = 0.02, H(3) = 9.62, *post hoc* Mann-Whitney U test S1–mPFC: p = 0.002 and S1-HPC: p = 0.01), probably due to the small sample sizes.

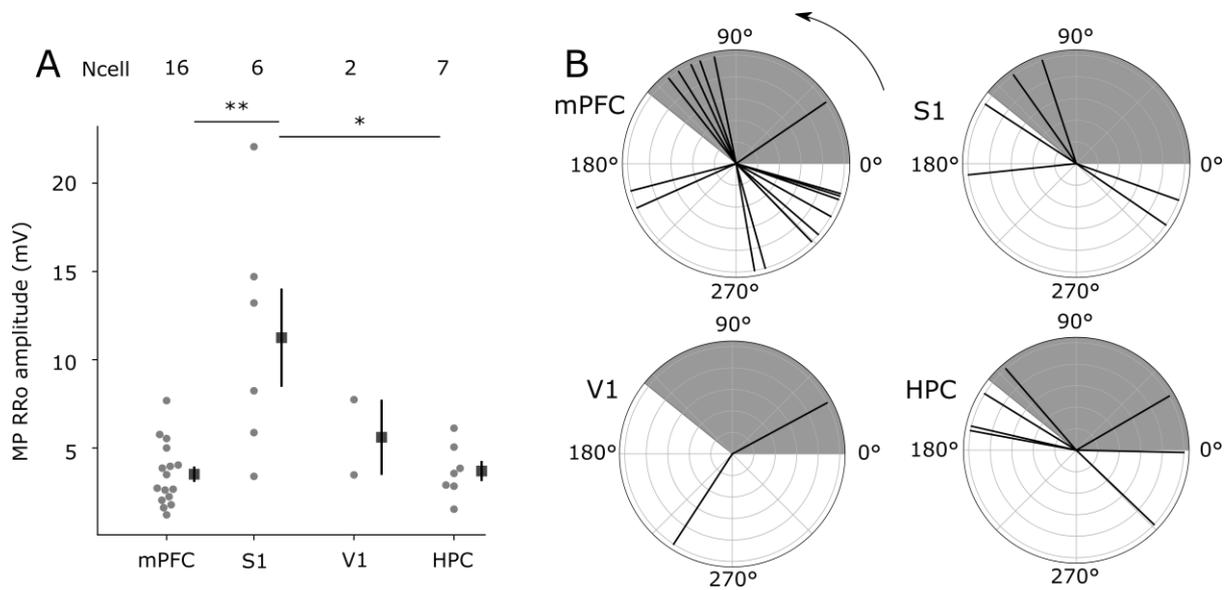

**Figure 3 : MP RRo phase and amplitude characteristics. Only recordings at basal MP are considered.** A. MP RRo amplitude across structures. Grey points represent MP RRo amplitude of individual cells, dark grey squares represent the mean by structure and black line represent SEM. Significant differences are observed (Kruskal-Wallis test, p = 0.02, H(3) = 9.62, $N_{mPFC}$ = 16 , $N_{S1}$ = 6, $N_{V1}$ = 2, $N_{HPC}$ = 7). *Post hoc* two-sided Mann-Whitney U test reveals amplitude of MP RRo is higher in S1 cells than in mPFC ones, p = 0.002. Only cells with MP RRo probability above 2.5% and strictly more than 5 cycles are considered. B MP RRo preferred respiratory-phase across structures. Preferred phases are presented in a trigonometric circle, time progressing is anticlockwise. Shaded and white areas, correspond respectively to inspirations and expirations.

To pursue, we characterized the respiratory phase of the MP RRo. As already suggested by the illustrative examples in Figure 1, MP RRo phases were not unique. Indeed, whereas MP exhibited maximal depolarized values around the transition phase between inspiration and expiration in S1 and HPC examples (Figure 1C2 and 1C4), MP of mPFC and V1 examples exhibited hyperpolarized values at this same respiratory phase (Figure 1C1 and 1C3). For a complete analysis, we extracted, for each cell, the phase of the most depolarized value of the median MP RRo (see Figure 1, panels C-bottom for examples of median MP RRo). Figure 3B displays for each structure, the phase for each cell. Globally these phases seem to be distributed according to two preferential values, between 130-170 degree and 315-340 degree, approximately corresponding respectively to I/E transition (at 140 degree) and the end of expiration before E/I transition point (E/I at 0 degree). For mPFC and S1, the majority of the vectors were distributed pretty much equally between the two phases, whereas for HPC 4/7 vectors were locked close to I/E transition, while the three others presented a different phase between end of expiration and beginning of inspiration.

Finally, we looked in details at the relationship between MP RRo and spiking discharge. Table 1 presents the different association's possibilities between MP (RRo or no RRo) and discharge (RR discharge and No-RR discharge). To construct this table, we considered the distribution of spikes as a function of respiratory phase (see Figure 1C, top panels) using previously described method used in olfactory bulb (Briffaud et al. 2012; Cenier et al. 2009) (see Material and Methods for details). For a given cell, we segregated spikes occurring during respiratory cycles presenting a RRo MP from those occurring during respiratory cycles presenting a no RRo MP. The two spike distributions were then classified as either respiration-related (RR) discharge or non-respiration-related (no RR) discharge. Table 1 summarizes this classification across cells for each structure. It shows the percentages of each association between spiking discharge patterns (columns) and respiration-triggered MP types (lines). In mPFC and S1, the RRo MP was mostly associated with the RR discharge pattern (80% for both structures) and conversely the No RRo MP was mostly associated with the No RR discharge pattern (71.4% and 85.7% respectively). Results about V1 follow the same trend with a smaller cell number.

Results about HPC did not show the same association profile. Indeed, only 25% of the MP RRo was associated with RR discharge pattern.

| Area | MP activity | | RR discharge | No RR discharge | Total cells |
|---|---|---|---|---|---|
| mPFC | RRo | $N_{cell}$ | 4 | 1 | 5 |
| | | % | 80% | 20% | |
| | No RRo | $N_{cell}$ | 6 | 15 | 21 |
| | | % | 28.57% | 71.43% | |
| S1 | RRo | $N_{cell}$ | 4 | 1 | 5 |
| | | % | 80% | 20% | |
| | No RRo | $N_{cell}$ | 1 | 6 | 7 |
| | | % | 14.29% | 85.71% | |
| V1 | RRo | $N_{cell}$ | 1 | 1 | 2 |
| | | % | 50% | 50% | |
| | No RRo | $N_{cell}$ | 1 | 2 | 3 |
| | | % | 33.33% | 66.66% | |
| HPC | RRo | $N_{cell}$ | 1 | 3 | 4 |
| | | % | 25% | 75% | |
| | No RRo | $N_{cell}$ | 3 | 5 | 8 |
| | | % | 37.5% | 62.5% | |

**Table 1: Neocortical areas show stronger coupling between MP and spiking than hippocampus.** For each cell, spikes where segregated depending if they occur during a MP RRo respiratory cycle or a no RRo respiratory cycles. Respiration triggered discharge histogram were computed for the two MP categories. Histograms are then classified as either RR discharge or no RR discharge. Results are presented for each area and each MP activity. Line "$N_{cell}$" represents the number of cell in each discharge class. Line "%" represents the percentage of cell in each discharge class relative to the total number of cells associated to a MP activity.

In summary, we showed that respiration-related membrane potential oscillations were larger for cells in S1. Their respiratory phases were mainly around I/E and E/I transitions for mPFC and S1 cells, while mainly around I/E transition for HPC cells. Moreover, respiratory coupling between MP and spiking discharge was much more frequent in mPFC and S1 (71.4% and 85.7% respectively) than in HPC (25%).

Faced with the transient nature of MP RRo and their variability in amplitude and phase, we asked if parameters such as cell intrinsic state or local network state could influence respiration entrainment of MP.

**Influence of intrinsic excitability and network activity on the respiratory modulation of membrane potential**

Our second aim was then to examine to what extent RRo MP could be influenced by *(i)* intracellular excitability, and *(ii)* by local field potential (LFP) state.

*(i) Effect of intracellular excitability.*

First, we studied the effect of excitability level on the probability for MP to be respiration-modulated. Briefly, we tested two hyperpolarization levels (Hyp1 and Hyp2) and one depolarized level (Dep) by progressive direct intracellular DC current injection. As described in materials and methods, we used 3 criteria (firing, MP value, and injected current) to classify these excitability states.

Figure 4A shows the proportions of respiration-entrained cells in each structure, as a function of internal excitability levels. mPFC and HPC displayed relatively stable proportions across the four different excitability levels (52.6 ± 5.1%, N = 4 excitability levels; and 82.0 ± 5.4%, N = 4 excitability levels; respectively). On the contrary, S1 dropped to 0% (0/4 cells) in Hyp2 and showed disparate proportions in other levels (Hyp1 = 20%, base = 60%, Dep = 25%). V1 showed stable proportion between base and Hyp1 levels (50%) and an increase to 80% at Hyp2 level. Thus, when looking at the whole population, excitability level did not clearly impact the proportion of respiration-entrained cells.

Then, we compared the percentages of MP RRo cycles for a same cell when it was hyperpolarized from base to Hyp1 level (Figure 4B). This cell-by-cell analysis was necessary because of the great variability of MP RRo cycles proportion at base level within a same structure (*cf.* Figure 2). This analysis indicated that, in mPFC and S1, the percentage of MP RRo cycles was considerably reduced when hyperpolarizing the cell (mPFC: Base: 5.7±2.9%, Hyper1: 1.6±1.1%, Wilcoxon test p = 0.02, Z = 0.0, N = 9 cells) (S1: Base: 15.7±7.1%, Hyper1: 1.0±0.6, Wilcoxon test p = 0.07, Z = 1.0, N = 4 cells).

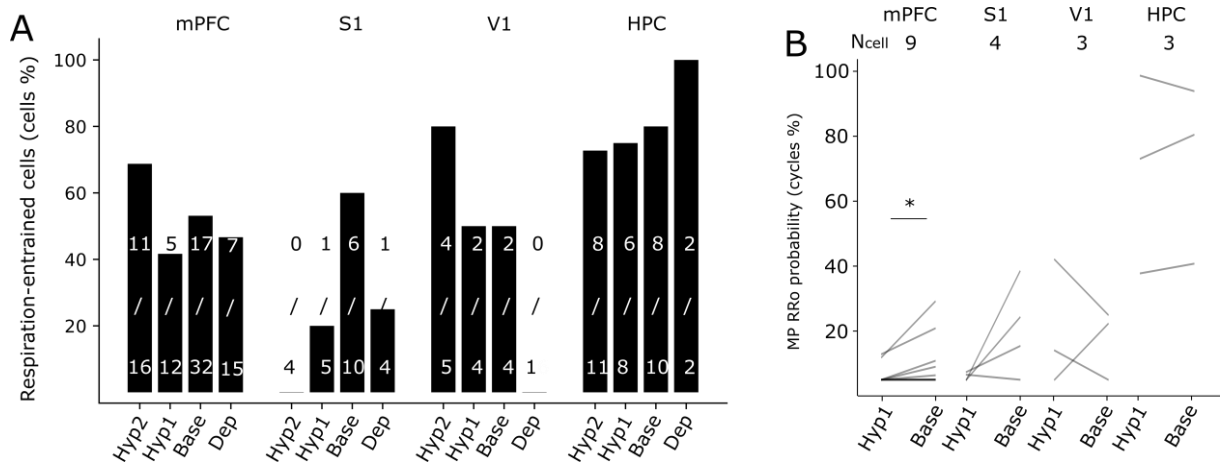

**Figure 4 : Internal excitability level has weak influence on percentage of respiration coupled cells on the whole population but decrease percentage of RRo MP cycles when cell by cell analysis.** A. Percentage of respiration-coupled cells across structures and internal excitability states. Numbers of cells appear in each black bar. B. Individual cell evolution of MP RRo probability across hyper1-to-base states. Only cells recorded in hyper1 and base conditions are presented. Recordings must be close in time (<7 min see methods). For mPFC, RRo probability is higher in hyper1 than in base (Wilcoxon paired test, p = 0.04, N =9).

These data showed that, even if internal excitability state had a weak influence on the proportion of respiration-entrained cells at the whole population level, moderate hyperpolarization of a cell, in PFC and S1, decreased its MP RRo cycle percentage.

*(ii) Effect of LFP state*.

Since it is now well established that LFP in these brain structures can be respiration-entrained (Girin et al. 2021; Tort et al. 2018), we then questioned how MP and LFP respiratory rhythms could interact. Raw data in Figure 5A1 shows an example where neither MP nor LFP were respiration-entrained. At this moment, MP and LFP signals appeared very different. Conversely, for the same cell, a few minutes later (Figure 5A2, after 2 min), respiratory frequency appeared simultaneously in MP and LFP signals. The similarity between both signals is highlighted when superimposing filtered respiratory signal (light grey traces on MP and LFP raw signals).

LFP RRo cycles were detected in the same way than MP RRo cycles, in order to be able to compare their temporal dynamics (see Methods). We found LFP RRo cycles in the four structures. Figure 5B displays the mean percentage of LFP RRo cycles per recording (grey dots) and the percentages per structure (blue squares). mPFC showed a much higher percentage of LFP RRo cycles (up to 74%, with a mean of 32.9±5.1%, N = 20 cells) than S1 (14.8±11.4%, N = 3 cells), V1 (15.4±9.4, N = 2 cells), or HPC (11.3±3.8%, N = 4 cells). Nonetheless, the differences were not significant, certainly due to low effectives (Kruskal Wallis test, p = 0.14, H(3) = 5.38). In the way that we did for the MP recordings, we quantified for each LFP recording the mean duration of respiration-entrained episodes. The mean number of consecutive LFP RRo cycles (Figure 5C) was significantly higher for mPFC (11.0±2.2, N = 20 cells) than for S1 (3.7±0.3, N = 3 cells), and HPC (4.2±0.4, N = 4 cells) (Kruskal Wallis test p

= 0.01, H(3) = 10.51, *post hoc* Mann-Whitney U test mPFC-S1: p = 0.01 and mPFC-HPC: p = 0.01)

Then, we observed to what extent LFP and MP respiration-entrainment could be correlated. To do so, we calculated the percentage MP RRo cycles when the LFP displayed RRo or not (Figure 5D). It clearly emerged that, independently of the structure considered, the MP had a higher probability to display RRo when the LFP displayed RRo. For mPFC, the increased MP RRo probability when LFP switched from no RRo to RRo was significant (Wilcoxon test p = 0.02, Z = 44, N = 21 cells). HPC exhibited the same tendency, and, to a lesser extent, S1 and V1 also. These results strongly suggest that LFP state could strongly influence MP. Reciprocally, we wondered if the probability that the LFP displayed RRo could be different when the MP displayed RRo or not (Figure 5E). Here we found that in mPFC, S1 and HPC, the LFP had a higher probability to display RRo when the MP displayed RRo, but this result was statistically significant only for mPFC (Wilcoxon test mPFC: p = 0.002, Z = 27, N = 21 cells). This confirms that, most of the time, the respiratory modulation of a cell MP is coherent with the respiratory modulation of its network environment. Note that the absence of statistical significance in regions others than mPFC is probably due to the small sample size.

To summarize, we observed that the respiratory modulation of MP activity is largely influenced by that of LFP.

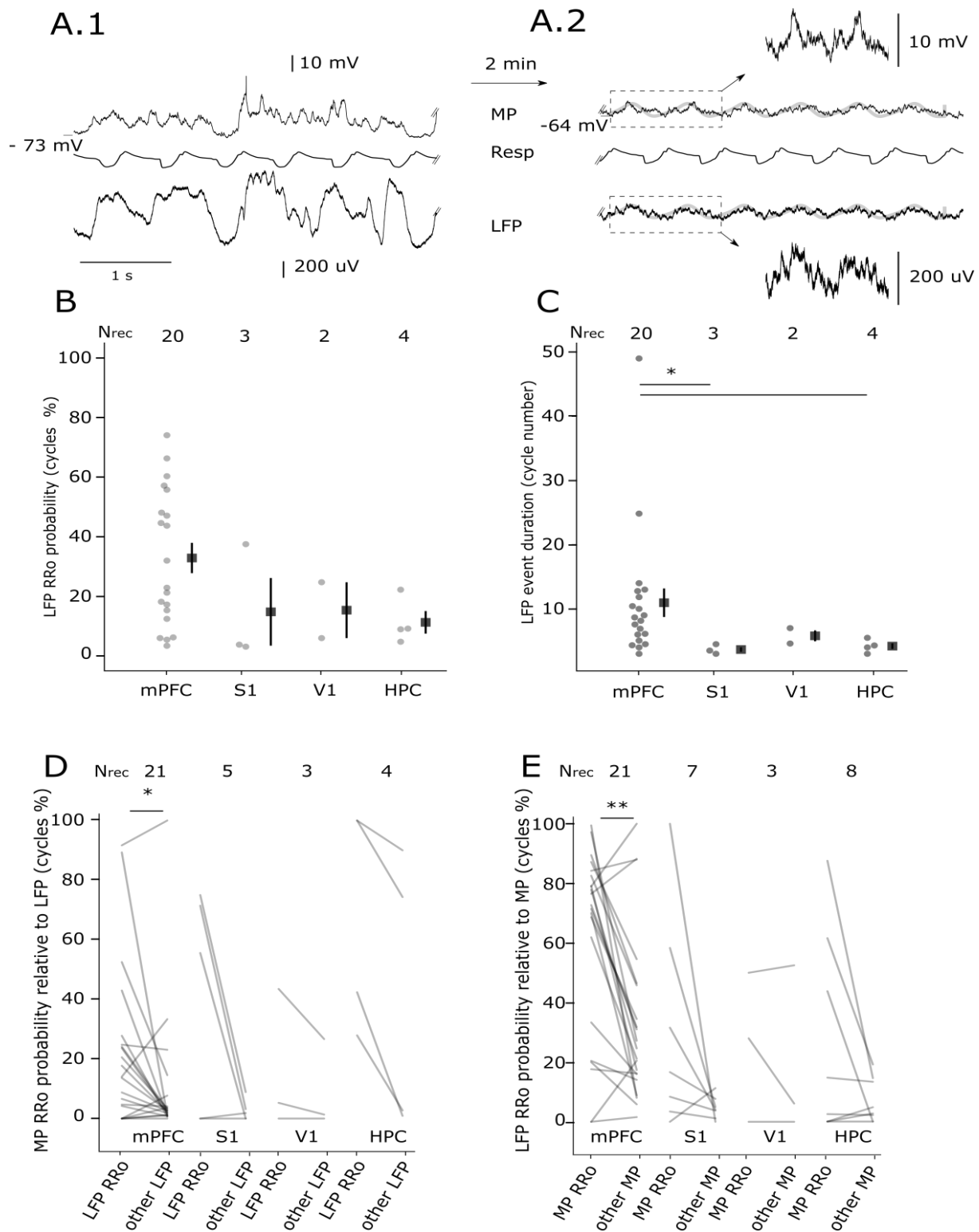

**Figure 5: MP RRo probability increases when RRos are concomitantly present in LFP.** Only recordings at basal MP are considered. A. MP (Top), respiratory signal (mid) and LFP (bottom) raw traces of the same cell when LFP and MP do not display RRo (A1) and two minutes later when LFP and MP display RRo (A2). On A2 panel, light grey traces show filtered

and time-shifted respiratory signal superimposed on MP and LFP traces. Time scale bar: 1s. B LFP RRo probability across structures. Grey points represent LFP RRo probability, dark grey squares represent the mean by structure and black line represent SEM. Only LFPs with RRo probability above 2.5% are considered. No significant difference is observed (Kruskal-Wallis test, $p = 0.14$, $H(3) = 5.38$, $N_{mPFC} = 20$, $N_{S1} = 3$, $N_{V1} = 2$, $N_{HPC} = 4$). ). C LFP RRo episode duration across structures. Grey points represent mean LFP RRo duration of individual traces, dark grey squares represent the mean by structure and black lines represent SEM. The same pool of traces as in B is used. Significant differences are observed (Kruskal-Wallis test, $p = 0.01$, $H(3) = 10.51$, $N_{mPFC} = 17$, $N_{S1} = 6$, $N_{V1} = 2$, $N_{HPC} = 8$). mPFC presents longer LFP RRo episodes than S1 (Mann-Whitney U test, $p = 0.01$) and HPC (Mann-Whitney U test, $p = 0.009$). D. Conditional probability of MP RRo given LFP dynamics. Columns "LFP RRo" represent the probability to observe MP RRo when LFP is also RRo. On the other hand, columns "other LFP" represent MP RRo probability when LFP is no RRo. Only cell recordings during which LFP presented both RRo and other oscillations were analyzed. In mPFC, MP RRo probability increases when LFP is RRo (Wilcoxon test, $p = 0.02$, $Z = 44$) D. Conditional probability of LFP RRo given MP dynamics. Columns "MP RRo" represent the probability to observe LFP RRo when MP is also RRo. On the other hand, columns "other MP" represent LFP RRo probability when MP is no RRo. Only cell presenting both MP RRo and other oscillations were analyzed. In mPFC, LFP RRo probability increases when MP is RRo (Wilcoxon test, $p = 0.002$, $Z=27$)

DISCUSSION

While many studies have recently evidenced a respiration-related activity in LFPs of many brain regions (Biskamp et al. 2017; Girin et al. 2021; Ito et al. 2014; Köszeghy et al. 2018; Liu et al. 2017; Tort et al. 2018; Yanovsky et al. 2014; Zhong et al. 2017), our study is the first to describe a respiratory-related oscillation in the MP activity in mPFC, S1, V1 and HPC. We defined a cell MP as respiratory modulated if at least 2.5% of its respiration-based cycles were respiration-locked. In such conditions, we showed for the first time a respiration-related oscillation of MP in a large proportion of cells (53.1% in mPFC, 60% in S1, 50% in V1, and 80% in HPC). Before us, very sparse and scarce data existed reporting MP oscillations coherent with LFP respiration-related oscillation ((Tort et al. 2018; Zhong et al. 2017): 1 cell recorded in parietal cortex; Yanovsky et al. 2014: 3 cells out 7 recorded in dentate gyrus after brainstem stimulation). Our study goes much further by comparing the probability of MP respiration-related oscillations in four different brain regions and by detailing how such oscillation could be influenced by excitability state or network activity.

We first evidenced that the proportions of cells expressing MP respiration-related oscillations were disparate between brain regions. In fact, 20% of recorded cells in S1 and mPFC showed about 20% of their MP cycles modulated by respiration, while more than 20% of HPC cells had about 75% of their cycles modulated, and no cell in V1 had more than 25% modulated cycles. The amplitude of MP RRo was larger in S1 than in other brain regions. Respiratory coupling between MP and spiking discharge was much more frequent in mPFC and S1 (100% and 80% respectively) than in HPC (25%). Interestingly, extracellularly recorded spiking activity in S1 has been previously described as rhythmically correlated with respiration (Ito et al., 2014). This stronger relation between respiration and S1 barrel cortex cells could have something to do with the fact that respiration binds different orofacial rhythms as whisking and sniffing particularly during olfactory search (Kleinfeld et al. 2014; Moore et al. 2013; Ranade et al. 2013).

We expected cells in mPFC to display more clearly respiration-related oscillations in their MP, given the strong modulation of LFP described in this region in the freely moving animal (Bagur et al. 2021; Dupin et al. 2019; Girin et al. 2021; Karalis et al. 2016). Indeed, neither the duration of MP respiration-modulated episodes nor the amplitude of MP RRo is more pronounced in mPFC than in the three other areas. The fact that anesthetics have voltage-dependent effects on resting MP (MacIver and Kendig 1991) could lower MP sensitivity to respiration. However, even in the awake head-fixed mouse, Köszeghy et al. 2018 reported that spiking discharges of neurons in mPFC are much less coupled to breathing than neurons in orbito-frontal cortex. This different probability of respiration-related oscillation between LFP and intracellular signals raises the question of a possible contamination of LFP signals by volume conduction coming from distant generators (Kajikawa and Schroeder 2011; Parabucki and Lampl 2017).

Oppositely, we showed a very clear respiration-related activity in MP of HPC cells (80% of cells expressed RRo and 20% of them displayed at least 75% of modulated cycles). This probably reflects the strong relation between olfactory and limbic structures. Very recently, Zhou et al. (2021) reported that human hippocampal connectivity is stronger with olfactory cortex than other sensory cortices and that olfactory-hippocampal connectivity oscillates with nasal breathing. Interestingly, several recordings of HPC neurons exhibiting MP RRo revealed that, when emitted in bursts, action potentials were locked to the I/E transition point of respiratory cycle while single action potentials were more loosely distributed over the respiratory cycle (data not shown). It is commonly admitted that bursts could facilitate synaptic transmission (Miles 1990) and induce synaptic plasticity, LTP or LTD, depending on the theta oscillatory cycle phase of the burst (Huerta and Lisman 1995). The fact that the MP of HPC cells could be paced by breathing could thus influence place cells coding notably during sniffing where respiratory and theta rhythms are in the same frequency range (Girin et al. 2021; Tort et al. 2018).

Although the small sample size of recorded cells in V1, 50% of them showed episodes of RRo MP. In these cells, the percentage of RRo MP did not exceed more than about 25% of total cycles. This confirms the observation that LFP in posterior areas are less susceptible to respiratory drive (Girin et al. 2021; Tort et al. 2018). It is nevertheless interesting to observe that, even in a region as posterior as V1 and so improbably connected to the olfactory input, respiration can affect subthreshold activity.

We previously showed that the MP of mitral/tufted cells, the principal neurons of the olfactory bulb, is strongly respiration-modulated (Briffaud et al. 2012). We cannot compare the percentages of modulated cells between the non-olfactory structures recorded here and olfactory bulb neurons because we did not use the same experimental paradigm, neither the same detection method nor data processing. However, it can be noted that hyperpolarization of mPFC and S1 cells appears to reduce the number of cells exhibiting RRo MP. On the contrary, in the olfactory bulb, hyperpolarization induced an increase in the proportion of respiratory modulated cells, which could suggest a reversal potential of RRo MP more hyperpolarized in non-olfactory structures than in olfactory bulb. It would be necessary to undertake a fine characterization of the different types of RRo that may exist (as in mitral/tufted cells) in order to be able to formulate hypotheses. However, we could postulate that a greater inhibition would be implicated in the oscillatory dynamics of these non-olfactory cells. Indeed, it has been shown that inhibitory interneurons appear to be more respiratory modulated than pyramidal neurons (Biskamp et al. 2017: mPFC; Karalis and Sirota 2018: mPFC and CA1 but not dentate gyrus). It would be of great interest to test the possibility that the consequence of a hyperpolarization results in rather an increase or a decrease of the respiratory modulation would come from the fact that the respiratory influence could rather impinge on inhibitory interneurons (as it could be in mPFC) or on a balanced excitatory-inhibitory network (as in OB). We previously showed the importance of excitation in respiratory modulation of spiking discharges and MP in the OB that mostly occurred through peripheral afferents on in mitral/tufted cells (Briffaud et al. 2012). This is confirmed by results of a preliminary experiment in mPFC showing that the amplitude of MP RRo decreases when the naris of the animal is blocked (data not shown).

We have been able to evidence RRo MP thanks to our detection method (see material and methods) which allowed us to detect even transient oscillations. Indeed, it is one of our major results that MP RRo episodes were temporary, often not exceeding five respiratory cycles in duration whatever the brain region, whereas the LFP RRo episodes were variable between regions, with duration up to 10 cycles in mPFC and between 3 and 6 cycles for other structures. This longer RR-LFP episode duration in mPFC could be due to its anatomic connections with olfactory bulb which put the mPFC much more often under the influence of olfactory pathway. Then, the respiratory rhythm may take over more often than other rhythms. Alternating populations of cells could transiently express respiratory modulation in their MP resulting in a longer duration of LFP respiratory modulation. In other structures, the competition between several other LFP rhythms (such as slow rhythm or theta) probably does not favor the emergence of this rhythm. Additionally, for all the structures, the sensitivity of each individual cell MP to LFP respiratory modulation likely differs according to its excitability state, limiting episodes of MP RRo to short durations. This could be supported by our observation that cells could show less RRo MP when hyperpolarized (see Figure 4B).

Although our study focused on the respiratory rhythm, the influence of other LFP rhythms on cellular activity (MP and discharge) should not be ignored. For examples, in few cases, we observed a spiking discharge that was RR while MP displayed no RRo. We could suppose that if the spiking discharge is influenced by a rhythm twice slower or twice faster than the respiratory rhythm, they will appear locked on respiratory cycle on spike histograms. We also observed more complicated patterns with several rhythms nested. For example, on the LFP spectrum analysis of Figure 1B3 we can distinguish a slow oscillation (around 1 Hz) which could compete with the respiratory rhythm.

Overall, our results evidenced a consequent proportion of RRo within electrophysiological signals, which could only have been revealed thanks to respiration co-recording. We evidenced a strong respiratory coupling between MP and spike discharge in PFC, S1 and V1 (and to a lesser extent in HPC). Furthermore, MP and LFP RRo dynamics showed strong covariations. Even if our experiments in anesthetized animal are purely descriptive, our results show that, at very different places in the brain, the MP of neurons can be depolarized and hyperpolarized according to the respiratory rhythm. These findings contribute to bring evidence that respiratory rhythm could be used as a common clock to set the dynamics of large-scale networks on a same slow rhythm.

ACKNOWLOEDGEMENTS

We would like like to thank Marc Thévenet and Belkacem Messaoudi for providing technical assistance.

REFERENCES USED